\documentstyle[12pt]{article}
\begin{document}
\begin{titlepage}
\pagestyle{empty}
\begin{center}
\large{{\bf Lattice Charge Overlap: Towards the Elastic Limit}}
\end{center}
\vspace{5 mm}
\begin{center}
Walter Wilcox
\end{center}
\begin{center}
{\it Department of
Physics, Baylor University, Waco, TX 76798}
\end{center}
\begin{center}
June, 1992
\end{center}
\normalsize
\setlength{\baselineskip}{10 mm}
\vspace{2 cm}

A numerical investigation of time-separated
charge overlap measurements is carried out for the pion
in the context of lattice QCD using smeared Wilson fermions.
The evolution of the charge distribution function is examined
and the expected asymptotic time behavior
$\sim e^{-(E_{q}-m_{\pi})t}$,
where $t$ represents the charge density relative time separation,
is clearly visible in the Fourier transform.
Values of the pion form factor are extracted
using point-to-smeared correlation functions and are seen to be
consistent with the expected monopole form from vector dominance.
The implications of these results for hadron structure
calculations is briefly discussed.
\end{titlepage}
\pagenumbering{arabic}

The use of charge overlap distribution functions to examine qualitative
features of lattice hadrons is now a well established
technique~[1-3]. However, there is a
need to go beyond qualitative measurements
to investigate whether such methods are capable of extracting
physical observables in the electromagnetic sector.
For this reason the Fourier transform of the
distribution functions for $\pi$ and $\rho$ mesons
were examined in Refs.~\cite{me2} and \cite{me3}.
It was found that the equal-time functions contain a substantial
contamination of intermediate states, preventing
a reliable extraction of electric and magnetic form factors
at available lattice momenta.

Consider now the continuum Euclidean
time-separated charge distribution for zero-momentum pions:
\begin{equation}
Q_{44}^{du}({\vec q\,}^{2},t) \equiv
\int d^{3}r
e^{i{\vec x}\cdot{\vec q}}P_{44}^{du}(r,t),\label{1}
\end{equation}
where
\begin{equation}
P_{44}^{du}(r,t) \equiv \frac{1}{2m_{\pi}}
(\pi^{+}({\bf 0})|
T[-J_{4}^{d}(\vec{r},t)J_{4}^{u}(0)]|\pi^{+}({\bf 0})),
\label{1.5}
\end{equation}
$J_{4}^{u,d}$ are the $u$, $d$ flavor charge densities.
and \lq\,$T$\rq\, is time-ordering.
Fig.~1 shows a symbolic representation of the
measurement in Eq.~({\ref{1.5}).
Assuming $t>0$ and
inserting a complete set of states, this can
be shown to result in~\cite{foot1}
\begin{equation}
Q_{44}^{du}({\vec q\,}^{2},t) = \sum_{X}\frac{
(\pi^{+}({\bf 0})|-J_{4}^{d}(0)|X({\vec q}))(X({\vec q})|
(J_{4}^{u}(0)|\pi^{+}({\bf 0}))}
{4E_{X}m_{\pi}}e^{-(E_{X}-m_{\pi})t}.\label{2}
\end{equation}
In the large Euclidean time limit, the sum reduces to a
single term, and we find that
\begin{equation}
Q_{44}^{du}({\vec q\,}^{2},t)
\stackrel{t\,\gg 1}{\longrightarrow}\frac{(E_{q}+m_{\pi})^{2}}
{4E_{q}m_{\pi}}F_{\pi}^{2}(q^{2})\,e^{-(E_{q}-m_{\pi})t}, \label{3}
\end{equation}
where (working in the flavor SU(2) limit)
\begin{equation}
F_{\pi}(q^{2})\equiv\frac{\left(\pi^{+}({\bf 0})|(J_{4}^{u},-J_{4}^{d})
|\pi^{+}(\vec{q})\right)}{m_{\pi}+E_{q}}. \label{4}
\end{equation}

Thus, by separating the currents in time
it is in principle possible to damp out the intermediate
state contributions and to measure form factors
at arbitrary lattice momenta. On a finite space-time lattice,
however, achieving this elastic limit is problematical. The exponential
damping factor, $e^{-(E_{q}-m_{\pi})t}$, is not large and the
fixed time positions of the initial and final pion interpolating
fields limit the possible charge density separation times.
Nevertheless, this paper demonstrates that the elastic limit can
essentially be achieved at low momentum values
on present-sized lattices. This has important implications
for hadron structure function calculations on the lattice.

We begin by examining the evolution of these
distribution functions as the time separation, $t$, between the
charge densities $J_{4}^{d}({\vec x},t)$ and $J_{4}^{u}(0)$ increases.
On the lattice, we measure
\begin{equation}
{\cal Q}_{44}^{du}({\vec q},t_{2},t_{1},t) \equiv
\sum_{{\vec x}}
e^{i{\vec x}\cdot{\vec q}}{\cal P}_{44}^{du}({\vec x},t_{2},t_{1},t),
\label{5}
\end{equation}
\begin{equation}
{\cal P}_{44}^{du}({\vec r},t,t_{1},t_{2}) \equiv
\frac{\langle {\rm vac}|T[-\phi^{sm.}(t_{2})
\sum_{{\vec x}}J_{4}^{d}({\vec x}+{\vec r},t_{1})
J_{4}^{u}({\vec x},t)\phi^{\dagger}({\vec 0},0)]|{\rm vac}\rangle}
{\langle {\rm vac}|T[\phi^{sm.}(t_{2})
\phi^{\dagger}({\vec 0},0)]|{\rm vac}\rangle},
\label{6}
\end{equation}
where we are now using lattice states and operators,
and the fields $\phi^{sm.}(t_{2})$ and $\phi^{\dagger}({\vec 0},0)$
are smeared and point pion interpolating fields, respectively.
We use the exactly conserved lattice charge densities
(which are non-local in time)
for $J_{4}^{d}({\vec x}+{\vec r},t_{1})$,
$J_{4}^{u}({\vec x},t)$.
Note that carrying out the sum on ${\vec x}$ in (\ref{6})
for all ${\vec r}$ takes
a nontrivial amount of computer time, scaling
as $N_{s}^{2}$, where $N_{s}$ is the number of
spatial points in the lattice.
One may show that for large time separations ($t_{2}-t_{1}$, $t_{1}-t$,
$t\,\gg 1$) and in the continuum limit,
\begin{equation}
{\cal Q}_{44}^{du}({\vec q},t,t_{1},t_{2}) \rightarrow
Q_{44}^{du}({\vec q\,}^{2},t_{1}-t),\,\,
{\cal P}_{44}^{du}({\vec r},t,t_{1},t_{2}) \rightarrow
P_{44}^{du}(r,t_{1}-t).
\label{7}
\end{equation}

The numerical results for this study are on $12$
quenched $\beta=6.0$ configurations
of a $16^{3}\times 24$ lattice at $\kappa=.154$ (the
largest $\kappa$ value studied in
Refs.~\cite{me3} and \cite{nucleon}.)
The pion interpolating sources were located at
time slices $4$ and $21$ and the
time-separated current densities
were positioned as symmetrically possible
in time between these sources.
Two quark propagators per configuration,
with origins at the positions of the interpolating sources,
were necessary to reconstruct these relative overlap functions.
Charge density self-contractions have been
neglected in forming these quantities.

Fig.~2 shows a ${\rm log}_{10}$ plot of the
spatial evolution of the charge overlap distribution
function, ${\cal P}_{44}^{du}$,
as the relative time separation between the $d,
u$ current densities is increased.
It is seen that the distribution
becomes flatter as the relative time is increased;
there is also a substantial change in the shape.
Points on the periodic lattice
boundary are given by the filled-in squares,
which are seen to have their values raised.
Note that these functions have been calculated using point-to-smeared
correlation functions; the quarks are smeared over the entire
volume of the lattice at time slice $21$ using the
lattice Coulomb gauge\cite{foot2} to produce zero-momentum
quark and hadron fields.

Fig.~3 represents a ${\rm log}_{10}$-plot of
the Fourier transform of these distribution
functions, ${\cal Q}_{44}^{du}$,
at the two lowest lattice spatial momentum values, $|{\vec q}|=\pi/8$,
$\sqrt{2}(\pi/8)$ as a function of relative time separation between
current densities. The solid lines shown in this figure come from
the expected asymptotic exponential falloff specified by Eq.~(\ref{3}),
using the (smeared) $\kappa=.154$
data from Table 1 ($m_{\pi}=.369, m_{\rho}=.46$) of
Ref.~\cite{nucleon}, assuming the vector dominance form for the pion
form factor: $F_{\pi}(q^{2})=1/(1+q^{2}/m_{\rho}^{2})$.
(We also assume the continuum relation $E_{q}=(m_{\pi}^{2}
+ {\vec q\,}^{2})^{1/2}$.) Actually shown
in this figure are results for both point-to-smeared ($\Diamond$)
as well as smeared-to-smeared ($\Box$) correlation functions.
In all cases, the expected functional dependence $\sim
e^{-(E_{q}-m_{\pi})t}$ is present, indicating that by
time step $7$ or $8$ single exponential behavior has emerged.
This is similar to the number of time steps needed in
hadron spectrum calculations.
This behavior is remarkable because although
we are damping out intermediate states as $t$ increases,
we are also moving closer to
possible contaminations from the
fixed interpolating fields at either
time end. In fact, we do not see any indications
of such contamination in the data.
These correlation functions are also unusual because the asymptotic
approach is from below, indicating damping of
negative terms in the $d, u$ correlation function.
Ref.~\cite{me3} indicates that these intermediate states,
at least in the continuum limit, are primarily positive G-parity
states\cite{foot3}.

Although the point-to-smeared and smeared-to-smeared
results exhibit essentially overlapping
error bars, the smeared-to-smeared values are
systematically low compared to the
point-to-smeared results, indicating a slight dependence on the
form of the interpolating field used. However, the results
of Ref.~\cite{me2} indicate that such a dependence decreases as
$\kappa\rightarrow\kappa_{critical}$, that is, as the physical
regime is approached.

In looking back at Fig.~2, it is clear
that the spatial distributions,
${\cal P}_{44}^{du}({\vec r})$, have substantial
contaminations near the periodic lattice
boundaries from the
surrounding image sources . Why is it then that we
have not attempted to do image corrections
on the Fourier transform, ${\cal Q}_{44}^{du}({\vec q})$, of
these distributions? Consider an infinite
spatial grid and associate the value
of a function, $F({\vec x})$, with
each point ${\vec x}\in \{ X\}$ of the grid.
Then, by doing a spatial translation on
points outside the given primitive cell (of size $L^{3}$)
to equivalent points, associated with
the same phase factor $e^{i{\vec q}\cdot{\vec x}}$,
inside the cell ${\vec x}\in \{ x\}$, it can be shown that
\begin{equation}
\sum_{{\vec x}\in \{ X\}}
e^{i{\vec q}\cdot{\vec x}}F({\vec x})=
\sum_{{\vec x}\in \{ x\}}
e^{i{\vec q}\cdot{\vec x}}{\hat F}({\vec x}),
\label{8}
\end{equation}
where for ${\vec x}\in \{ x\}$
\begin{equation}
{\hat F}({\vec x})\equiv\sum_{{\vec n}}F({\vec x}+{\vec n}L).
\label{9}
\end{equation}
(${\vec n}=(n_{x},n_{y},n_{z})$, where  $n_{x,y,z}$ are integers.)
The terms with ${\vec n}\neq 0$ are the image contributions.
Eq.~(\ref{8}) says that the discrete
Fourier transform of $F({\vec x})$ on the
infinite grid is the same as the
transform of ${\hat F}({\vec x})$
in the primitive cell for ${\vec q}$ values allowed
in the original primitive cell.
In our case this means there are no image corrections
to ${\cal Q}_{44}^{du}({\vec q})$, assuming
the distributions ${\cal P}_{44}^{du}({\vec r})$ are
strictly periodic.

By fitting time steps 7 through 10 of Fig.~3 with a
single exponential of known slope and removing the
kinematical factor in Eq.~(\ref{3}), a measurement of
$F_{\pi}(q^{2})$ can be made. The results for the two
calculated ${\vec q\,}^{2}$ values using point-to-smeared
correlation functions are shown in Fig.~4.
(The error bars in Figs.~3 and 4 were determined from first
and second-order single elimination jackknifes, respectively.)
The values found are consistent within errors with vector
dominance, shown as the solid line. Also shown in this figure
are the results from a previous three-point-function simulation
of the pion form factor\cite{pion}. Comparison of the two
types of results shows that the error bars of these different
techniques are of the same order of magnitude for similar numbers
of configurations. (Note the different $\beta$
values and lattice sizes of the two simulations, however.)

The significance of these results for hadron
structure calculations goes beyond the specific findings
discussed here for form factors. The ability to reach the
elastic limit for hadron four-point-functions
is crucial in attempting to perform
direct simulations of hadron structure functions. Up to the present,
attention has been focused on the moments of such functions,
which are given by the operator product expansion.
These expansions
are based upon separation of the short-distance physics,
calculated perturbatively, from the long-distance part,
evaluated on the lattice
(or by other techniques) in the form of certain
operator expectation values. The presence on the lattice, in the
continuum limit, of all appropriate
physical scales, combined with the ability to make contact
with the elastic limit of charge overlap four-point-functions,
demonstrates that direct simulations of structure functions are
feasible with current technology, at least at low lattice momenta.

What additional quantities need to be calculated in order to
initiate such a program? Clearly, the overlap
function considered here, $Q^{du}_{44}({\vec q\,}^{2},t)$ is a
sub-dominant quantity for structure functions; the
short-distance physics will mainly be
contained in the same-flavor Fourier transform:
$Q^{uu}_{44}({\vec q\,}^{2},t)$ or
$Q^{dd}_{44}({\vec q\,}^{2},t)$.
Calculating this piece involves putting both electromagnetic
current densities on the same quark line. This
type of zero-momentum
four-point function is difficult to simulate
on the lattice, but can be obtained by combining
fixed-source and quark-smearing techniques
and is currently under investigation. In addition, the
the current-current overlap distribution function,
$(\pi({\bf{0}})|\sum_{l}J_{l}({\vec r},t)J_{l}(0)|\pi({\bf{0}}))$,
for both flavor-diagonal and nondiagonal currents is necessary
in order to complete measurements of the two standard
structure functions, $W_{1}$ and $W_{2}$. Of course it is
the proton, not the pion, which is of most
phenomenological interest. However, there is no barrier to
applying these techniques to the proton as well, as long
the lattices used are able to contain all the necessary length
scales and the elastic limit in all these
various four-point-functions can be demonstrated.

To summarize, this paper shows that the elastic limit
of $d,u$ charge overlap distribution functions,
in the context of the pion, can essentially
be achieved at low momentum values
on present-sized lattices. We do not claim that all the systematics
of such measurements are yet completely
understood, but the message that this limit is very
close is extremely encouraging
to attempts to provide direct
calculations of structure functions from the
fundamental QCD Lagrangian.
\vspace{1 cm}

This work was partially supported by the National Center for Supercomputing
Applications and utilized the NCSA CRAY~2 system at the University of
Illinois at Urbana-Champaign.

\pagebreak
\Large
{\bf Figure Captions}
\normalsize
\vspace{7 mm}
\begin{description}
\item[Fig.~1:] Symbolic representation of the $d,u$ charge
overlap measurement. The pion sources, shown as circles, are
at fixed time positions in the lattice.
\item[Fig.~2:] Evolution of the lattice
$d$, $u$ spatial charge density correlation
function, ${\cal P}_{44}^{du}$, for pions
as the time separation, $t_{1}-t$, between
$J_{4}^{d}({\vec r},t_{1})$, $J_{4}^{u}({\vec 0},t)$
operators varies, as a function of non-equivalent lattice $r$ values.
(a) $t_{1}-t=0$. (b) $t_{1}-t=3$.
(c) $t_{1}-t=6$. (d) $t_{1}-t=10$.
Filled-in squares identify points on the periodic lattice boundary.
\item[Fig.~3:] ${\rm Log}_{10}$ plot of the
Fourier transform of the spatial
charge density correlation function, ${\cal Q}^{du}_{44}$,
as a function of time separation $t_{1}-t$ between
$J_{4}^{d}({\vec r},t_{1})$, $J_{4}^{u}({\vec 0},t)$
operators. The upper points give the $|{\vec q}|=\pi/8$
results, the lower points correspond to
$|{\vec q}|=\sqrt{2}(\pi/8)$. Results are given
for both point-to-smeared correlation functions ($\Diamond$)
as well as smeared-to-smeared correlation functions ($\Box$).
\item[Fig.~4:] Pion form factor, $F_{\pi}$, as a
function of $q^{2}/m_{\rho}^{2}$. The solid line is the
monopole form from vector dominance.
The data shown are the present results from point-to-smeared
correlation functions ($\Diamond$) and
previous results on the pion form factor from Ref.~\cite{pion}
($\Box$).
\end{description}

\begin{thebibliography}{9}
\bibitem{foot1} In a high energy inelastic context,
the discrete sum $\displaystyle{\sum_{X}}$
in Eq.~(\ref{1}) would be replaced by an integral
over the appropriate density of states.
\bibitem{foot2} These are the same Coulomb gauge-fixed configurations
used in Refs.~\cite{me2} and \cite{nucleon}. No
attempt was made to eliminate possible Coulomb gauge Gribov copies.
\bibitem{foot3} Note that an
alternate, momentum-smeared (no ${\vec x}$ sum in
Eq.~(\ref{6})) $d, u$ correlation
function discussed in Ref.~\cite{me3}
never exhibits the crucial exponential behavior $\sim
e^{-(E_{q}-m_{\pi})t}$ as the charge densities are
separated in time $t$; it is no longer
considered a viable measurement
of the continuum distribution $P_{44}^{du}(r,t)$ by the present author.
\end{thebibliography}
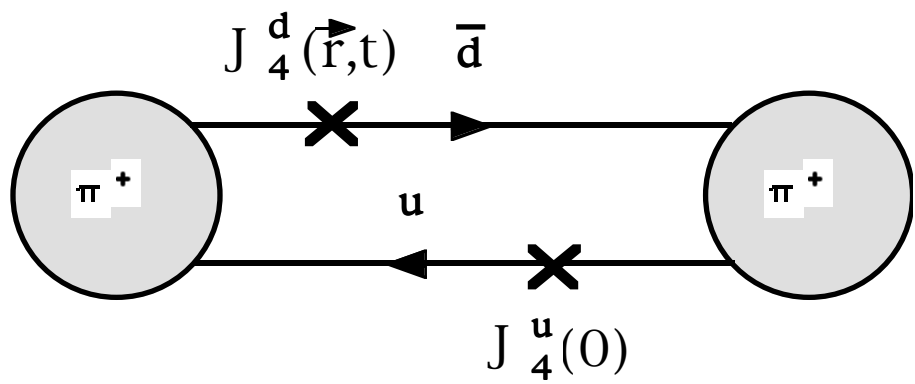
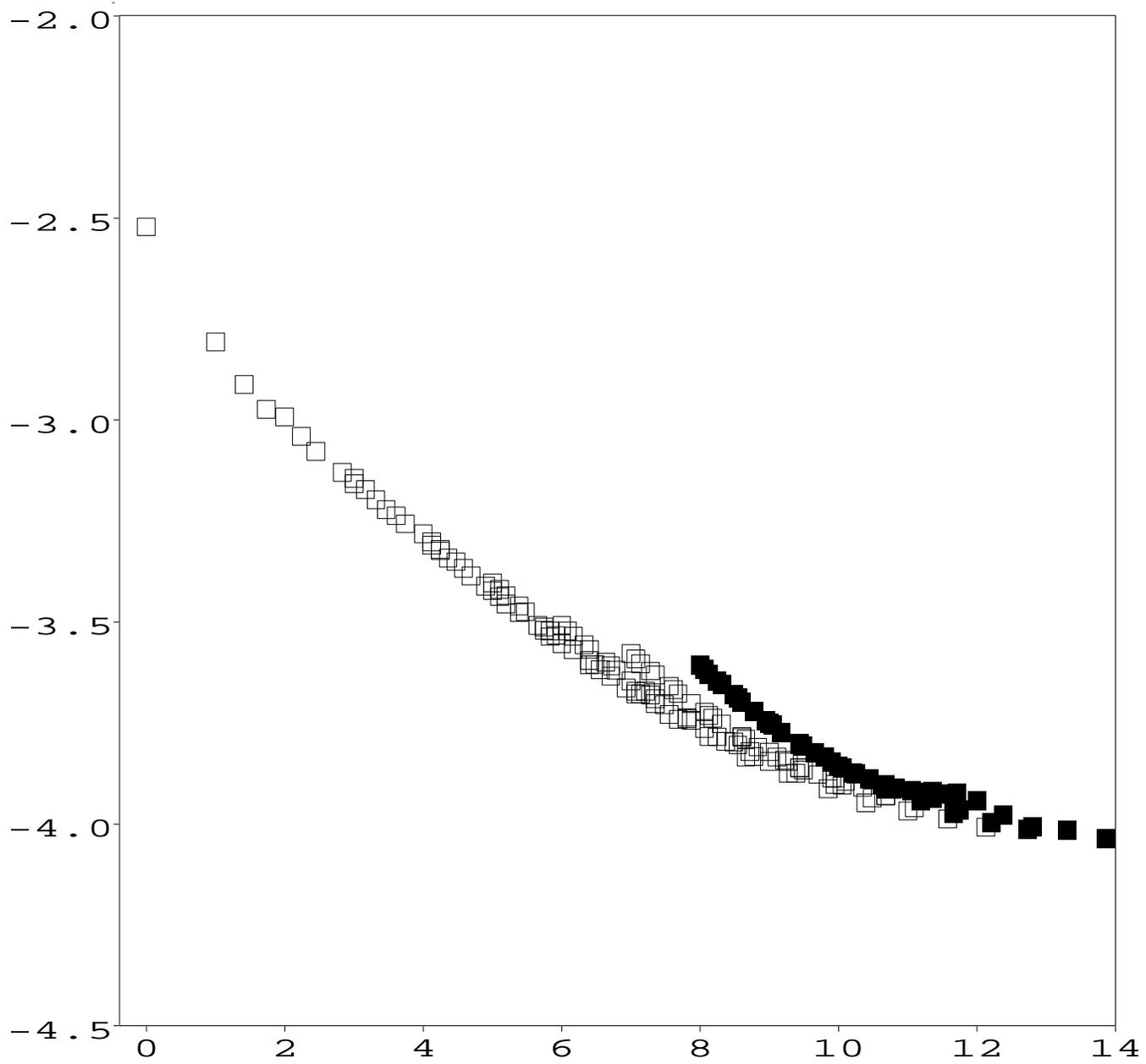
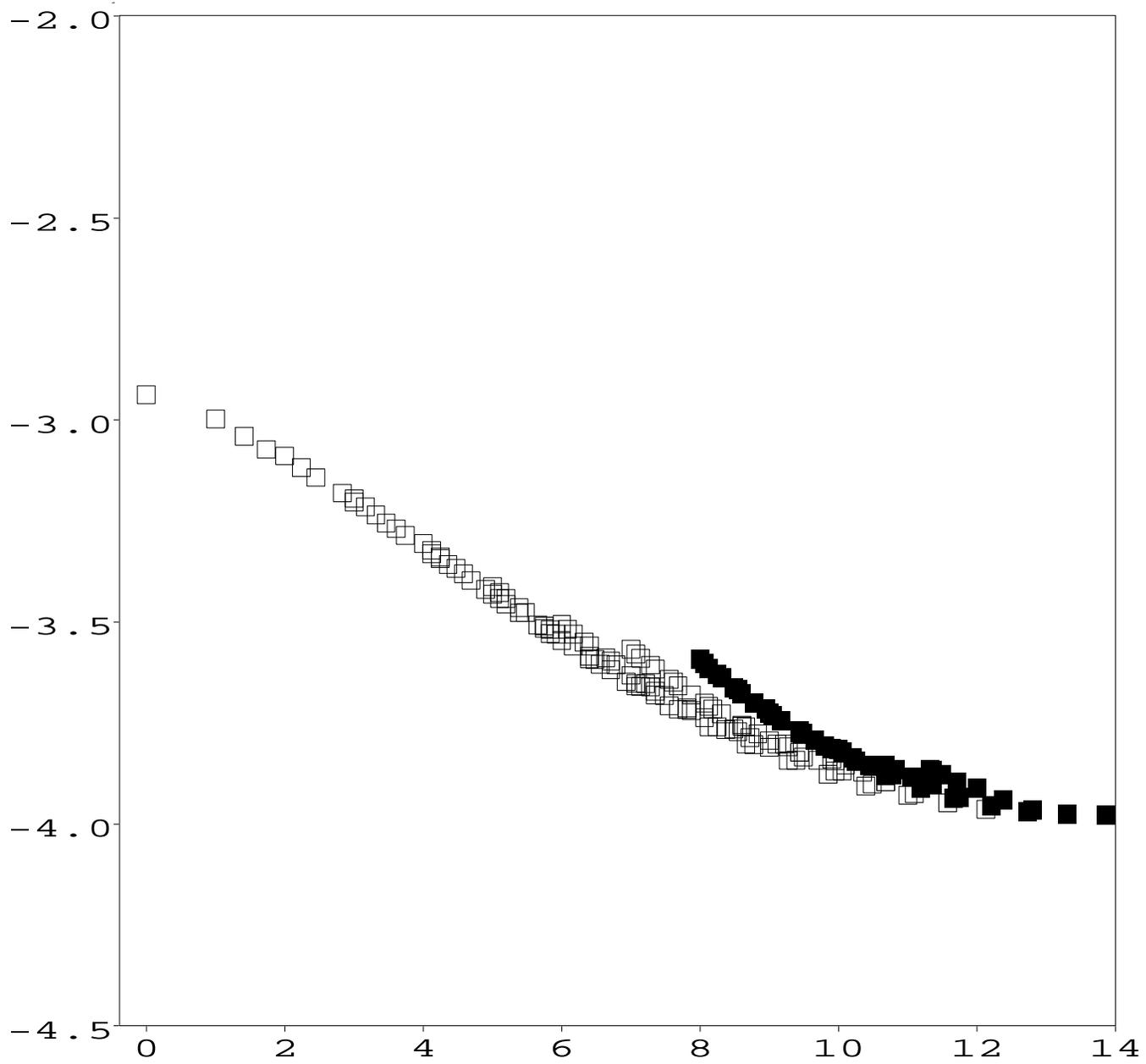
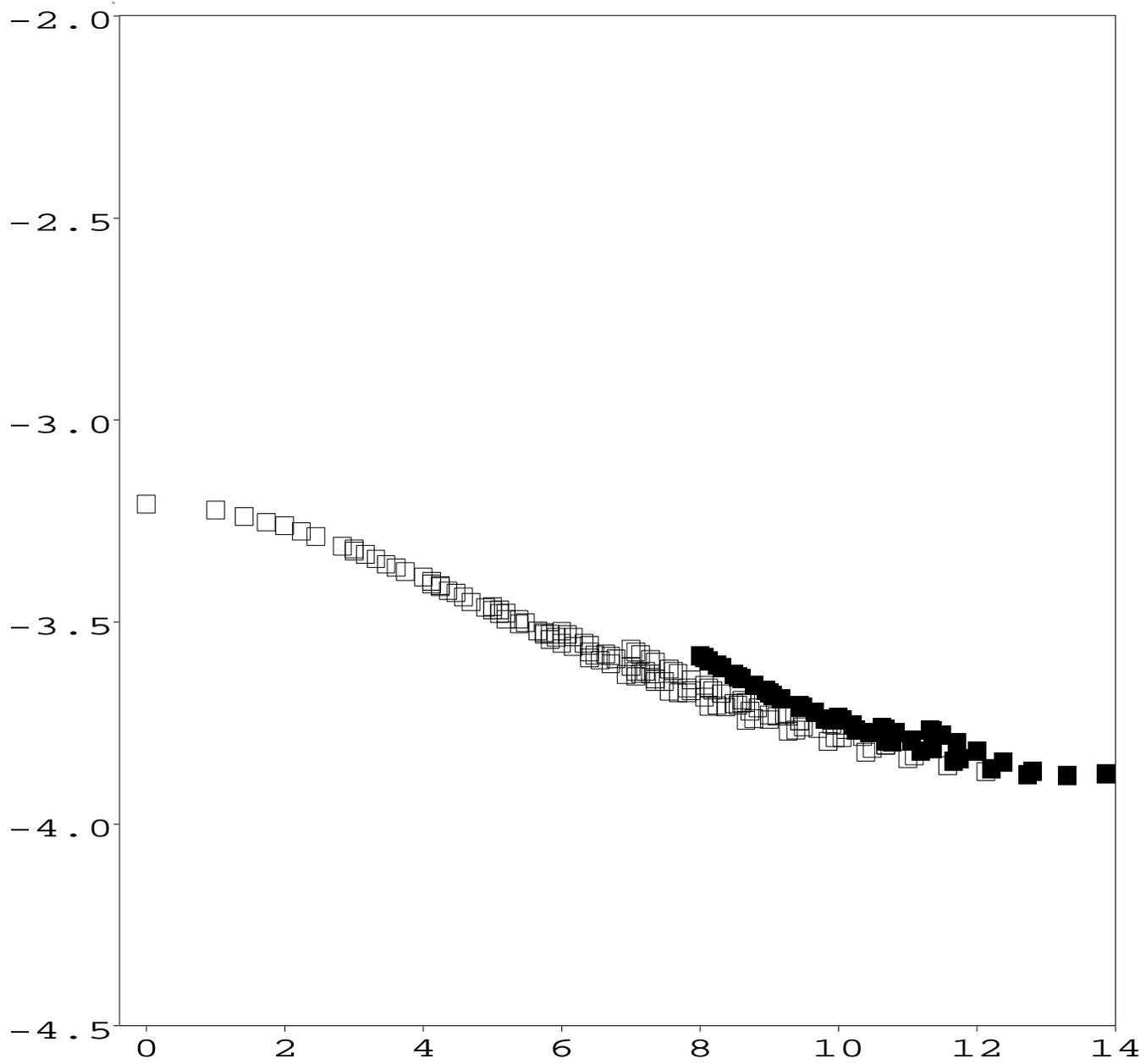
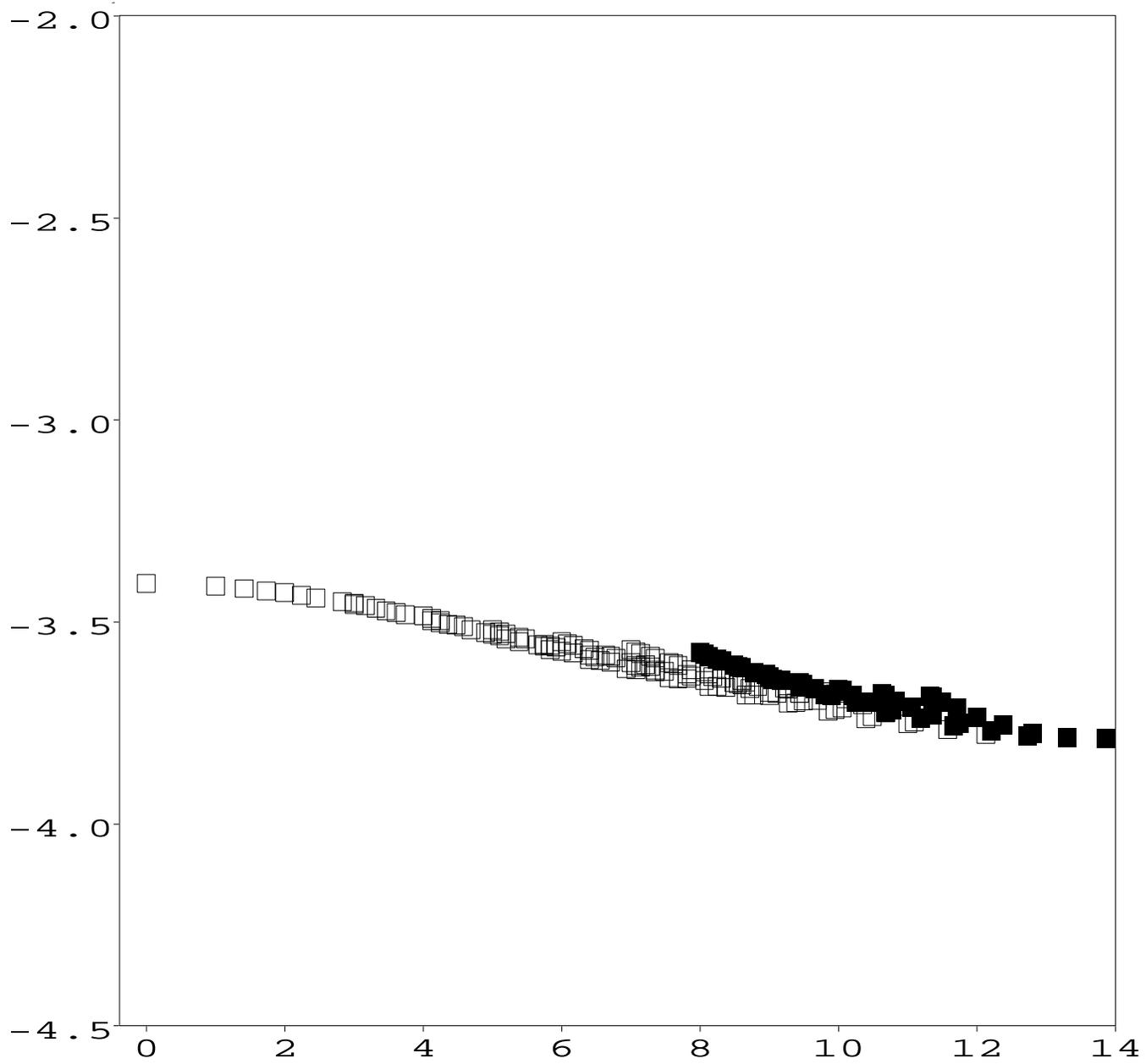
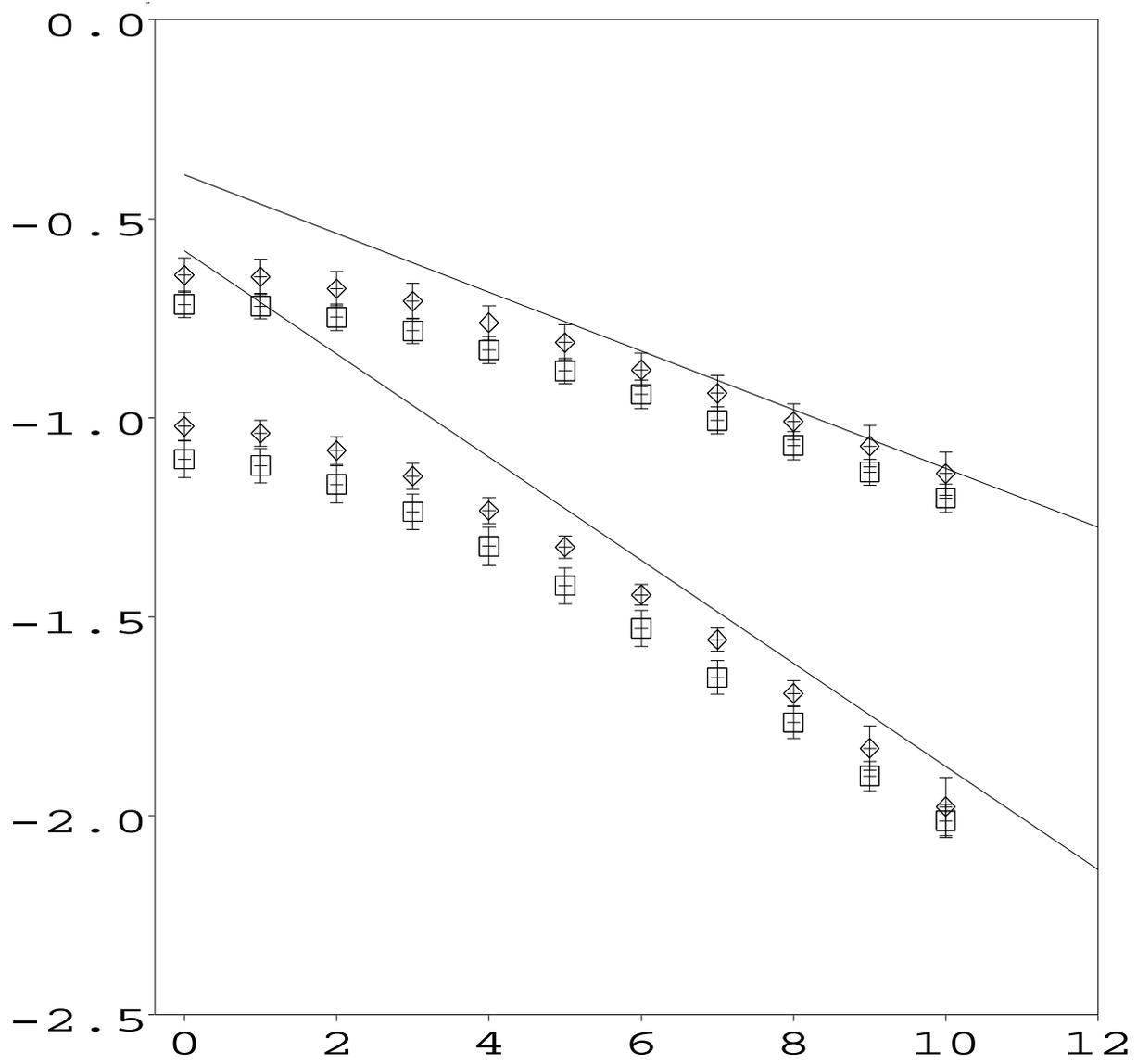
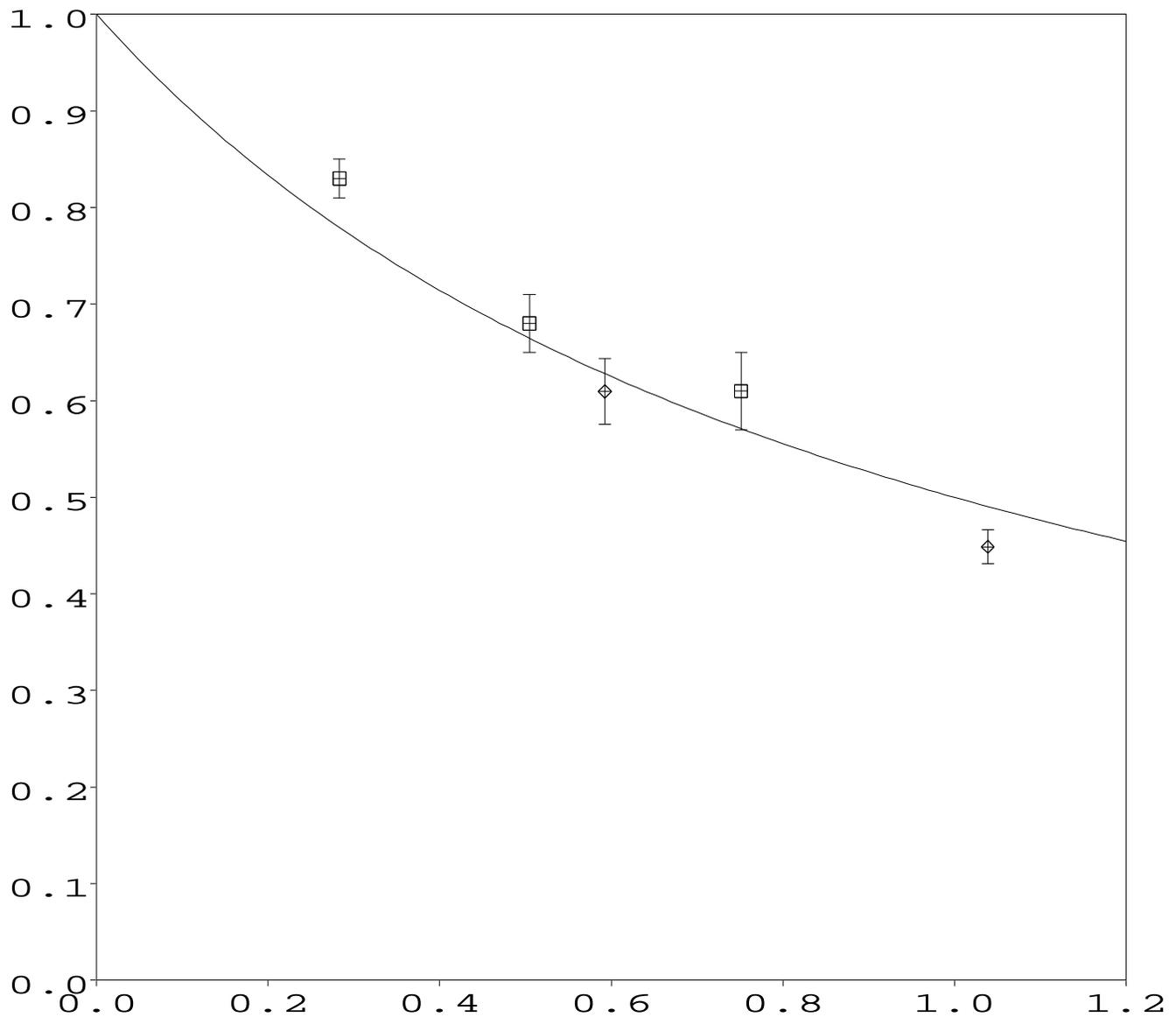
\end{document}